\documentclass[a4paper,10pt]{article}
\usepackage[english]{babel}

\input{Qcircuit}

\title{Quantum gambling using mesoscopic ring qubits}

\author{Ireneusz Paku\l a \footnote{email: ipakula@wp.pl}\\ University of Silesia \\ Institute of Physics \\ ul. Uniwersytecka 4 \\ 40-007 Katowice, Poland}

\begin{document}

\maketitle

\begin{abstract}
Quantum Game Theory provides us with new tools for practising games and some other risk related enterprices like, for example, gambling. The two party gambling protocol presented by Goldenberg {\it et al} is one of the simplest yet still hard to implement applications of Quantum Game Theory. We propose potential physical realisation of the quantum gambling protocol with use of three mesoscopic ring qubits. We point out problems in implementation of such game.
\end{abstract}

\section{Introduction}
A major development in applying quantum mechanical formalism to 
various fields has been made during the last few years. Quantum 
extensions of Game Theory \cite{EWL,PS3}, Economy \cite{PS1,PS2,PS4,PS5}, as well as diverse approaches 
to Quantum Information Theory have been found and currently are 
being explored; the notion of quantum artificial intelligence \cite{MPS} has also been proposed. We present a possible physical implementation of the two party gambling protocol described by Goldenberg {\it et al} \cite{G1}, which can be the smallest commercially useful application of a quantum computing device.

\section{The Game}
Two players, Alice and Bob, share one two-qubit state prepared by Alice (Bob can operate on just one of the qubits only). Let us name the qubits A and B after the players' names. If Bob finds his qubit in state $\ket{1}$, he wins one monetary unit. If he chooses to measure both qubits and finds them in a state other than $(\ket{10}+\ket{01})/\sqrt{2}$ he wins $R$ units. In other cases he loses one unit to Alice.

The optimal strategy for Alice is to prepare qubits exactly in state $(\ket{10}+\ket{01})/\sqrt{2}$. Bob's strategy is to use an auxiliary qubit C, perform a certain rotation and then measure his qubit B. If he finds his qubit in state $\ket{0}$, he verifies the initial state preparation. Depending on $R$, his expected maximal gain $G$ for $R\to\infty$ is described by:
\[
G=-\sqrt{\frac{2}{R}},
\]
when rotating his BC qubits around the angle $\theta$ given below:
\[
\sin^2 \theta=\sqrt{\frac{1}{2R}},
\]
regardless of Alice's strategy. The graphical representation of the gambling protocol is given by the following quantum circuit:
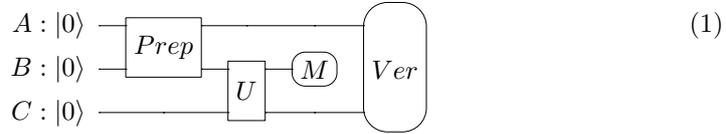
\begin{equation}
\Qcircuit @C=1em @R=0.7em {
\lstick{A:\ket{0}} & \multigate{1}{Prep} & \qw & \qw & \multimeasure{2}{Ver} \\
\lstick{B:\ket{0}} & \ghost{Prep} & \multigate{1}{U} & \bgate{M} \\
\lstick{C:\ket{0}} & \qw & \ghost{U} & \qw & \ghost{Ver}
}
\end{equation}
where {$Prep$} is Alice's state preparation
\[
\ket{00}\to \alpha\ket{00}+\beta\ket{01}+\gamma\ket{10}+\delta\ket{11},
\]
the unitary transformation {$U$}
\[
U=\left[\begin{array}{cccc}
1 & 0 & 0 & 0 \\
0 & \cos\theta & -\sin\theta & 0 \\
0 & \sin\theta & \cos\theta & 0 \\
0 & 0 & 0 & 1
\end{array}\right]
\]
is Bob's rotation of BC qubits, $M$ is the measurement of Bob's qubit and {$Ver$} is the verification measurement -- projection of AC qubits onto state
\[ \sqrt{\frac{\sin^2\theta}{1+\sin^2\theta}}\ket{01}+\sqrt{\frac{1}{1+\sin^2\theta}}\ket{10}. \]

\section{Implementation}
Mesoscopic ring qubits, among other applications like fidelity measurement \cite{Les1}, seem to be a promising way to give us scalable hardware able to do quantum computing and gaming. The Hamiltonian in the case of a non-superconducting ring with an energy barrier
\[
H=\sum_{m\neq n}[E_n\ket{n}\bra{n}-\frac{1}{2}\hbar (\omega_{mn}\ket{n}\bra{m}+\omega_{mn}\ket{m}\bra{n})]
\]
in certain conditions \cite{Zip1} reduces to the general Hamiltonian describing persistent current flux qubits \cite{Lup1,Les1}:
\[
H=-\frac{1}{2}(\epsilon\sigma_z+\Delta\sigma_x)
\]
thus opening an interesting and feasible way to implement solid state qubit. Three level qubits \cite{K1,KP1} (using the third auxiliary level for quantum gate operations) are currently under development as well.

Realisation of such a game with use of mesoscopic ring qubits is related with solving of two main problems. First of all, players need to prepare Alice's initial state and then perform Bob's rotation. Those are two qubit unitary transformations resulting in three qubit entangled state. To do this we need a way to entangle states of two rings -- possibly by interaction with microwave cavity \cite{KP1}. As any unitary transformation can be approximated using elementary quantum gates, the desired operations can also be accomplished by applying a sequence of qubit-flips, phase flips, Hadamard gates and CNOTs -- CNOT, which is the most complicated gate of those, has been realised by Yamamoto {\it et al.} \cite{Y1} in the context of charge qubits and shown by Plourde {\it et al.} \cite{Plo1} with use of the flux devices. Another way to realise those universal gates is described by Kulik \cite{K1}.

Measurement of a single qubit is just a current direction measurement which can be done using magnetic flux measurement methods, just like the one proposed by Lupascu {\it et al.} \cite{Lup1}, but we encounter a serious obstacle -- to perform quantum gambling two qubit measurement is needed (that is, projection on a certain two qubit state). How to perform such a measurement is still an open problem.

\section{Summary}
We presented a realisation of a quantum gambling game with use of one of the most promising nanotechnology devices -- mesoscopic rings acting as Aharonov-Bohm qubits. Commercially applicable tasks could be thus performed by system of just three coupled rings, if only forementioned problems were overcome.

\section*{Acknowledgements}
This paper has been supported by the {\bf Polish Ministry of Scientific Research and Information
Technology} under the (solicited) grant No {\bf PBZ-MIN-008/P03/2003}.

\end{document}